\documentclass[aps,prl,twocolumn,superscriptaddress,final]{revtex4-2}

\pdfoutput=1

\usepackage[colorlinks=true,allcolors=blue]{hyperref}
\usepackage{graphicx}
\usepackage{amsmath,amssymb,mathrsfs,array,longtable,delarray}
\usepackage{dcolumn}
\usepackage{bm}
\usepackage{textcomp} 

\begin{document}

\title{Magnetotransport Spectroscopy of Strongly Rashba-Split Hole Subbands \\
Reveals Many-Body Interactions}

\author{F. Sfigakis}
\thanks{These authors contributed equally to this work}
\affiliation{Institute for Quantum Computing, University of Waterloo, Waterloo N2L 3G1, Canada}
\affiliation{Department of Electrical and Computer Engineering, University of Waterloo, Waterloo N2L 3G1, Canada}
\affiliation{Department of Chemistry, University of Waterloo, Waterloo N2L 3G1, Canada}

\author{N. A. Cockton}
\thanks{These authors contributed equally to this work}
\affiliation{Department of Physics and Astronomy, University of Waterloo, Waterloo N2L 3G1, Canada}

\author{M. Korkusinski}
\thanks{These authors contributed equally to this work}
\affiliation{Emerging Technologies Division, National Research Council of Canada, Ottawa K1A 0R6, Canada}

\author{S. R. Harrigan}
\affiliation{Institute for Quantum Computing, University of Waterloo, Waterloo N2L 3G1, Canada}
\affiliation{Department of Physics and Astronomy, University of Waterloo, Waterloo N2L 3G1, Canada}
\affiliation{Waterloo Institute for Nanotechnology, University of Waterloo, Waterloo N2L 3G1, Canada}

\author{G. Nichols}
\affiliation{Institute for Quantum Computing, University of Waterloo, Waterloo N2L 3G1, Canada}
\affiliation{Department of Physics and Astronomy, University of Waterloo, Waterloo N2L 3G1, Canada}

\author{\\Z. D. Merino}
\affiliation{Institute for Quantum Computing, University of Waterloo, Waterloo N2L 3G1, Canada}
\affiliation{Department of Physics and Astronomy, University of Waterloo, Waterloo N2L 3G1, Canada}

\author{T. Zou}
\affiliation{Department of Physics and Astronomy, University of Waterloo, Waterloo N2L 3G1, Canada}
\affiliation{Waterloo Institute for Nanotechnology, University of Waterloo, Waterloo N2L 3G1, Canada}

\author{A. C. Coschizza}
\affiliation{Department of Physics and Astronomy, University of Waterloo, Waterloo N2L 3G1, Canada}
\affiliation{Waterloo Institute for Nanotechnology, University of Waterloo, Waterloo N2L 3G1, Canada}

\author{T. Joshi}
\affiliation{Institute for Quantum Computing, University of Waterloo, Waterloo N2L 3G1, Canada}
\affiliation{Department of Electrical and Computer Engineering, University of Waterloo, Waterloo N2L 3G1, Canada}

\author{A. Shetty}
\affiliation{Institute for Quantum Computing, University of Waterloo, Waterloo N2L 3G1, Canada}
\affiliation{Department of Chemistry, University of Waterloo, Waterloo N2L 3G1, Canada}

\author{M. C. Tam}
\affiliation{Department of Electrical and Computer Engineering, University of Waterloo, Waterloo N2L 3G1, Canada}
\affiliation{Waterloo Institute for Nanotechnology, University of Waterloo, Waterloo N2L 3G1, Canada}

\author{\\Z. R. Wasilewski}
\affiliation{Institute for Quantum Computing, University of Waterloo, Waterloo N2L 3G1, Canada}
\affiliation{Department of Electrical and Computer Engineering, University of Waterloo, Waterloo N2L 3G1, Canada}
\affiliation{Department of Physics and Astronomy, University of Waterloo, Waterloo N2L 3G1, Canada}
\affiliation{Waterloo Institute for Nanotechnology, University of Waterloo, Waterloo N2L 3G1, Canada}

\author{S. A. Studenikin}
\affiliation{Emerging Technologies Division, National Research Council of Canada, Ottawa K1A 0R6, Canada}

\author{D. G. Austing}
\affiliation{Emerging Technologies Division, National Research Council of Canada, Ottawa K1A 0R6, Canada}

\author{J. B. Kycia}
\email{jkycia@uwaterloo.ca}
\affiliation{Institute for Quantum Computing, University of Waterloo, Waterloo N2L 3G1, Canada}
\affiliation{Department of Physics and Astronomy, University of Waterloo, Waterloo N2L 3G1, Canada}
\affiliation{Waterloo Institute for Nanotechnology, University of Waterloo, Waterloo N2L 3G1, Canada}

\author{J. Baugh}
\email{baugh@uwaterloo.ca}
\affiliation{Institute for Quantum Computing, University of Waterloo, Waterloo N2L 3G1, Canada}
\affiliation{Department of Chemistry, University of Waterloo, Waterloo N2L 3G1, Canada}
\affiliation{Department of Physics and Astronomy, University of Waterloo, Waterloo N2L 3G1, Canada}
\affiliation{Waterloo Institute for Nanotechnology, University of Waterloo, Waterloo N2L 3G1, Canada}


\begin{abstract}
We report the results of magnetotransport experiments carried out on low-disorder 2D hole gases (2DHG) in the strongly correlated liquid regime, hosted in dopant-free (100) GaAs/AlGaAs single heterojunctions. Over a wide range of 2DHG densities (from $0.7 \times 10^{15}$/m$^2$ to $2 \times 10^{15}$/m$^2$), Fourier analysis of low-field (B~$<$~1~T) Shubnikov–de Haas oscillations reveals two spin-orbit-split heavy-hole (HH) subbands with distinct effective masses contributing to transport. Surprisingly, the lighter-mass HH subband exhibits a parabolic dispersion with Fermi wavevector below the anticrossing between the heavy-hole and light-hole subbands, while the heavier HH subband is non-parabolic throughout. Quantitative comparison with numerical calculations based on the Luttinger model reveals that both effective masses are enhanced by a common factor ($\approx$\,2.3), which we attribute to many-body interactions. This common scaling factor has a very weak dependence on the 2DHG density, likely due to band hybridization. Our measured hole masses are compared with published cyclotron resonance and magnetotransport values. We propose a cohesive framework reconciling the long-standing three-way discrepancy between Luttinger theory, magnetotransport, and cyclotron resonance measurements of density-dependent effective masses in partially spin-orbit-polarized heavy-hole systems in GaAs.
\end{abstract}

\maketitle

Many aspects of semiconductor transport in two-dimensional holes gases (2DHGs) involving strong spin orbit interactions (SOI) remain not well understood, regardless of material system. In stark contrast to the conduction band, there are no analytical expressions for (and no consensus on) the most elementary 2DHG transport properties, such as dispersion relations, the density of states, or the effective mass. Critically, band hybridization (a.k.a light-hole heavy-hole mixing) and band non-parabolicity invalidate most of the analytical tools used to calculate parameters from transport experiments such as, e.g., the Fermi energy $E_\textsc{f}$, Fermi wavevector $k_\textsc{f}$, hole-hole interaction strength, Landau level mixing, and Landau level spacing. This severely limits feedback from experiments to theory.

\begin{figure}[t]
    \includegraphics[width=1.0\columnwidth]{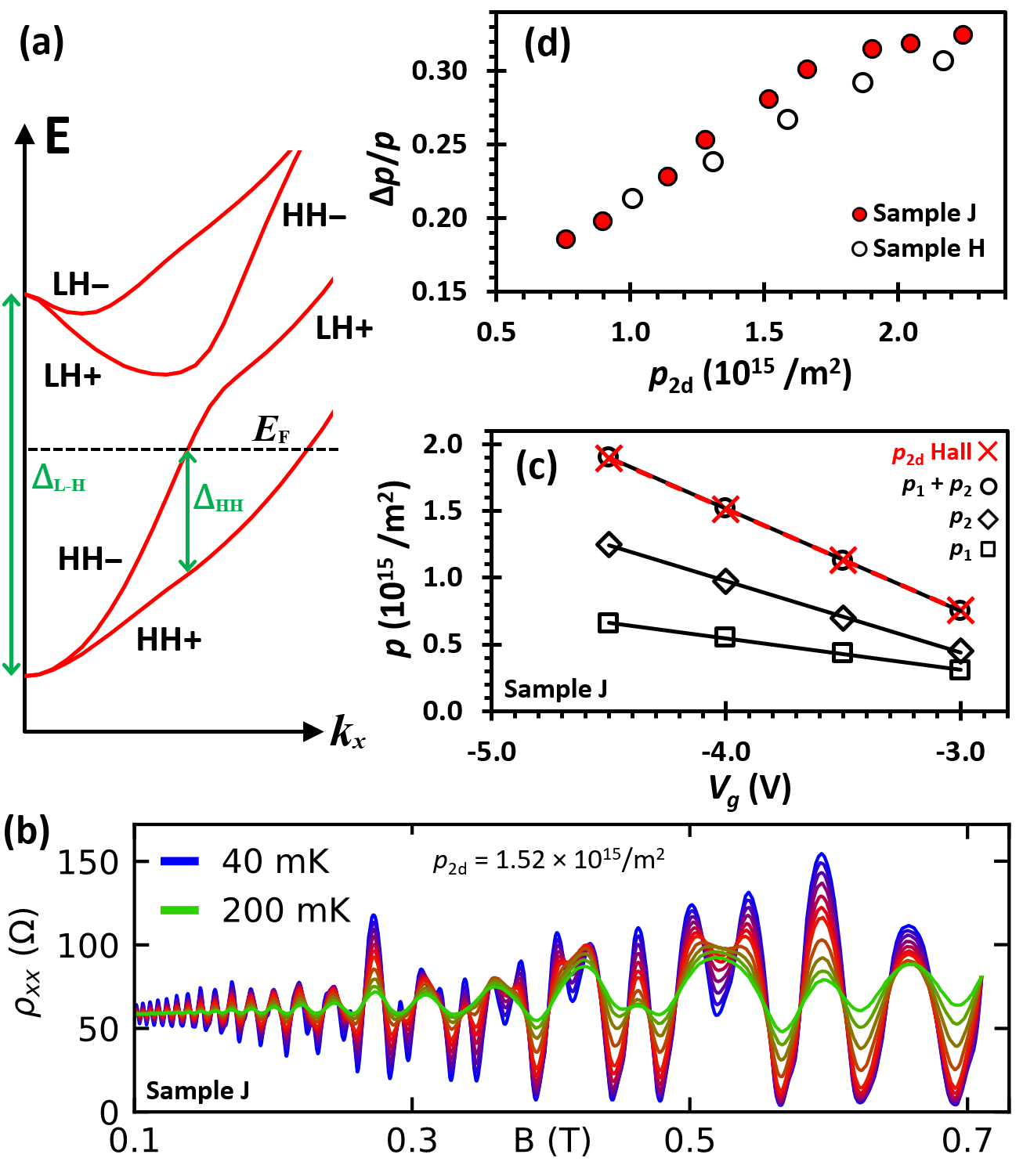}
    \caption{(a) Diagram of the dispersion relations for spin-orbit polarized LH and HH subbands \cite{companionPRB2025}, illustrating the anticrossing of HH$-$/LH$+$ in the presence of LH/HH hybridization. (b) Shubnikov-de Haas (SdH) oscillations at a fixed 2DHG density. (c) Comparison between carrier densities extracted from SdH oscillations and the Hall effect. The quantities $p_1$ (squares), $p_2$ (triangles), and $p_1 + p_2$ (circles) were obtained from the Fourier transform analysis of SdH oscillations. Cross symbols represent $p_{\text{2d}}$ determined from the low-field Hall effect. Solid (dashed) lines are linear least-squares fits to the FT (Hall) data. (d) Spin-orbit polarization $\Delta p/p$ versus 2DHG density.  }
    \label{fig:Basics}
\end{figure}

In principle, knowledge of the effective mass $m^*$ as a function of $E_\textsc{f}$ would give direct insight into the dispersion relation $E(k)$ in the valence band (where $k$ is the crystal momentum), if $m^*$ were to be measured in transport experiments at equilibrium with methods that do not explicitly rely on a parabolic dispersion. One must take into account that effective masses in the valence band are anisotropic in space (a consequence of holes' p-orbital wavefunction), that the relevant mass for 2DHGs is the in-plane $m^*_\parallel$ (not the out-of-plane $m^*_\perp$ obtained from optical studies), and that the effective mass changes with carrier density (a consequence of band non-parabolicity). Furthermore, in the presence of spin orbit interactions, structural inversion asymmetry (SIA; a.k.a. Rashba SOI) lifts spin degeneracy at zero magnetic field in every valence subband, and the two spin species from each subband acquire different $m^*$. We refer to this regime as spin-orbit polarization, to distinguish it from conventional spin polarization driven by Zeeman splitting. Figure~\ref{fig:Basics}(a) illustrates a partially spin-orbit polarized 2DHG at Fermi energy $E_\textsc{f}$ in zero magnetic field ($B=0$), where only the lowest heavy hole (HH) subband is populated and the lowest light hole (LH) subband is unpopulated. Henceforward, we will label heavy holes with the lower carrier density $p_1$ and lighter effective mass $m_1$ to the HH$-$ subband, and heavy holes with the higher carrier density $p_2$ and heavier effective mass $m_2$ to the HH$+$ subband.

In this Letter, we report the effective masses $m_1$ and $m_2$ of the spin-orbit split HH$-$ and HH$+$ subbands in dopant-free (100) GaAs/AlGaAs single heterojunctions, over a wide range of 2DHG densities $p_{2d}$, without assuming \textit{a priori} any particular subband structure. Surprisingly, we find that the HH$-$ dispersion is nearly parabolic at Fermi wavectors below the LH/HH anticrossing. Thus if the dispersion relation $E(k)$ of HH$-$ contains any linear or cubic term(s) in $k$, they are too small to be measured, which is incompatible with widely used theory models for Rashba interactions (e.g., \cite{winkler2000rashba}). Furthermore, a parabolic HH$-$ dispersion enables the use of conventional analytical tools for transport experiments, albeit in modified form. For example, the modified analytical expression for the Fermi energy slightly differs from the one widely used for 2DHGs, which can overestimate $E_\textsc{f}$ by 50\% even if using the correct effective mass. These findings also extend to 2DHGs in asymmetric quantum wells for $p_{2d}$~$>$~$1.3 \times 10^{15}$/m$^2$, but not to symmetric quantum wells. We propose a new experimental signature for observing the anticrossing between HH$-$ and LH$+$, based on spin-orbit polarization. We demonstrate quantitative agreement between our density-dependent effective masses and those reported from cyclotron resonance and transport experiments, previously taken at single 2DHG densities. Relative to predictions from the Luttinger model, we observe both $m_1$ and $m_2$ to be enhanced by the same scaling factor, and argue this scaling is due to renormalization from hole-hole interactions. This renormalization appears to be independent of the 2DHG density over the range investigated, which could be due to band hybridization (heavy-hole/light-hole mixing) in a total angular momentum $J=3/2$ system. We propose a framework that reconciles the long-standing three-way discrepancy between Luttinger theory, magnetotransport, and cyclotron resonance measurements of effective masses of spin–orbit-split heavy-hole systems in GaAs/AlGaAs single heterojunctions.

Gated Hall bars in the form of heterostructure-insulator-gate field effect transistors (HIGFET) \cite{Harrell1999fabrication, Willett2006simple, Wendy2010distinguishing, pan2011impact, ChenJCH2012fabrication, croxall2013demonstration, WangDQ2013influence, Sebastian2016gating, croxall2019orientation} were fabricated from GaAs/Al$_{0.3}$Ga$_{0.7}$As (100) single heterojunctions, grown by molecular beam epitaxy (MBE) without any intentional doping. A gate, insulated by 300~nm of SiO$_2$, tuned the carrier density with voltage $V_g$. Other fabrication details are described elsewhere \cite{companionPRB2025}. All data presented here come from two devices, samples~J and H from different wafers. Four-terminal measurements of the longitudinal resistivity $\rho_{xx}$ were performed at low temperature in a $^3$He/$^4$He dilution refrigerator with an electron temperature of $\sim\,$40~mK, using standard low-frequency lock-in techniques with a small ac excitation current of 2~nA at 17.3~Hz. The observation of fractional quantum Hall states (not shown; see Fig.~3 in \cite{companionPRB2025}) is consistent with the samples' high mobilities 60$-$84~m$^2$/Vs at $(1.4-1.8)\times 10^{15}$/m$^2$.

At low magnetic fields, Figure~\ref{fig:Basics}(b) shows a complex pattern in $\rho_{xx}(B)$, due to the interference of Shubnikov-de Haas (SdH) oscillations from two separate ladders of Landau levels with different effective masses, belonging to the HH$+$ and HH$-$ subbands. Conventional transport techniques for characterizing the effective mass and quantum scattering times are therefore not applicable. This problem is circumvented by performing all data analysis in Fourier space \cite{nichele2014spin, nichelethesis, companionPRB2025}. To measure $m_2$, the magnetic field must be below the onset of the quantum Hall effect (i.e., before the SdH oscillation minima reach $\rho_{xx}(B)=0$~\textohm), and the experimental conditions must otherwise satisfy $(\frac{\hbar eB}{m_{1,2}}-h\Gamma) > k_B T $, where $k_B$ is the Boltzmann constant, $\Gamma$ is the quantum scattering rate, and $h$ ($\hbar$) is the (reduced) Planck constant. To remove the uncertainty associated with any possible dependence of sample parameters on $B$, we simultaneously measure $m_1$ and $m_2$ over the same magnetic field interval.

Figure~\ref{fig:Basics}(c) confirms there is no parallel conduction in sample J because $p_{\text{Hall}} = p_{2d} = p_1 + p_2$; this is also the case in sample~H (not shown). This is independently confirmed by the minima of SdH oscillations from the integer/fractional quantum Hall effects reaching $\rho_{xx}=0$~\textohm ~for up to $B$~=~15~T (not shown).

Strong SIA causes a large energy difference $\Delta_{\textsc{hh}}$ between the HH$+$ and HH$-$ subbands [see Fig.\,\ref{fig:Basics}(a)]. Without means to directly measure $\Delta_{\textsc{hh}}$, we use the spin-orbit polarization $\Delta p/p = (p_2 - p_1)/(p_2 + p_1)$ as a proxy for $\Delta_{\textsc{hh}}$. Figure~\ref{fig:Basics}(d) shows $\Delta p/p$ increasingly monotonically as a function of $p_{2d}$. A more negative gate voltage produces a steeper confining potential for the 2DHG and, consequently, a stronger SIA with increasing $p_{2d}$ \cite{lu1998tunable, grbic2008strong, nichele2014spin}. This observed low-field ($B<0.7$~T) spin-orbit polarization is the second largest ever reported in GaAs single heterojunctions or quantum wells \cite{stormer1983energy, Cole1997, habib2004spin, grbic2004, nichele2014spin, yuan2009landau, Rendell2022,lu1998tunable}.

\begin{figure}[t]
    \includegraphics[width=1.0\columnwidth]{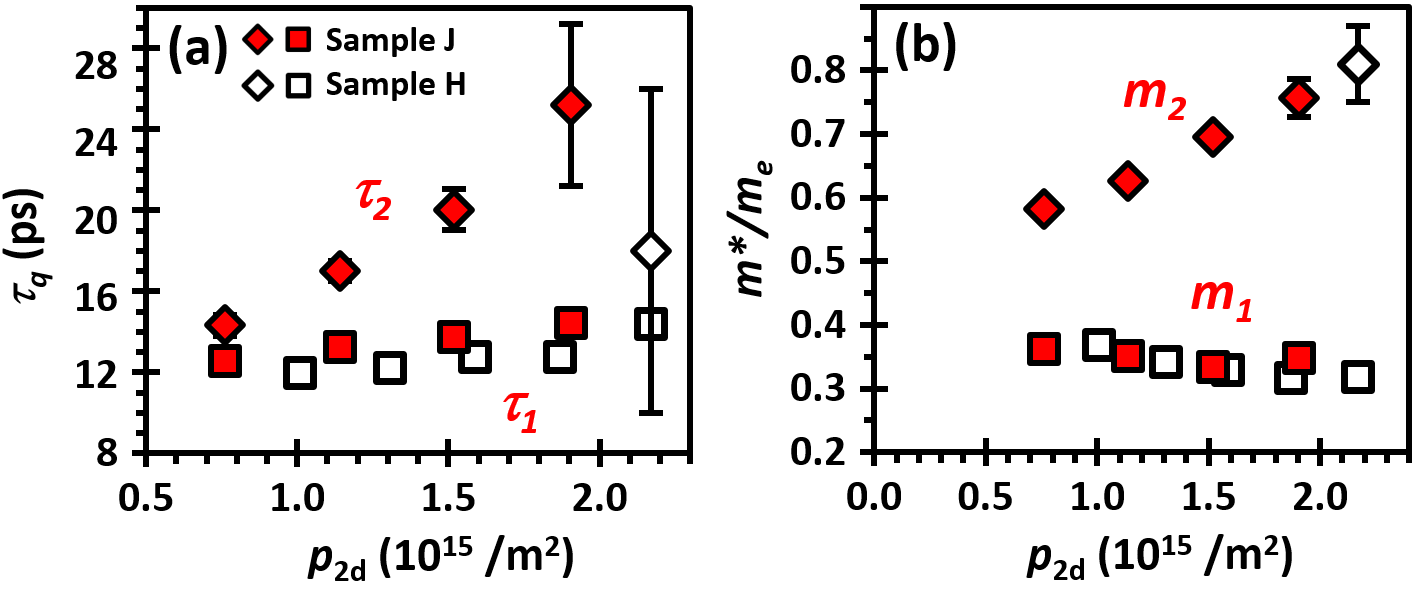}
    \caption{Fit values of: (a) the quantum scattering times ($\tau_q$) and (b) the effective massess ($m_1, m_2$), obtained from the Fourier analysis of SdH oscillations at different 2DHG densities in sample J (full symbols) and sample H (empty symbols), for HH$-$ (squares) and HH$+$ (diamonds). }
    \label{fig:FFT}
\end{figure}

Figure~\ref{fig:FFT} plots the fit parameters resulting from the Fourier analysis on the SdH oscillations. Its most striking feature is the near-independence of $m_1$ on the 2DHG density. Thus the dispersion relation of the HH$-$ subband is essentially parabolic ($E=\hbar^2 k^2/2m^*$), over the 2DHG density range covered. The median value of the HH$-$ effective mass is $\widetilde{m_1} = (0.35\pm0.01)m_e$ in sample~J and $\widetilde{m_1} = (0.34\pm0.03)m_e$ in sample~H. These values agree very well with each other, and with past reports of the effective mass measured at a single 2DHG density, obtained from low-field SdH oscillations in (100) GaAs/AlGaAs single heterojunctions in the same experimental conditions that we investigated: $m_1= (0.34\pm0.01)m_e$ at $p_{2d} = 3.0\times 10^{15}$/m$^2$ \cite{grbic2004}, and $m_1= (0.36\pm0.03)m_e$ at $p_{2d} = 5.0\times 10^{15}$/m$^2$ \cite{stormer1983energy}. Collectively, these results suggest that the parabolic dispersion regime for the HH$-$ subband may be valid for a very wide range of 2DHG densities in single heterojunctions, from $p_{2d} = 0.7\times 10^{15}$/m$^2$ to $5\times 10^{15}$/m$^2$.

Accurate Fermi energy $E_{\textsc{f}}$ of the 2D system and Fermi wavevector $k_{\textsc{f}}$ for HH$-$ can now be directly calculated from experiments using:
\begin{equation}
k_{\textsc{f}} = \sqrt{4\pi p_1}
\qquad\text{and}\qquad
E_{\textsc{f}} = 2\pi\hbar^2p_1/m_1.
\label{Eq:E_F(p1)}
\end{equation}
The above is strictly only valid for $E_{\textsc{f}}$ or $k_{\textsc{f}}$ values below the HH$-$/LH$+$ anticrossing point. Note the usual factor of two for spin degeneracy is missing, since the system is already spin-orbit polarized at $B=0$. The HH$-$ subband carrier density ($p_1$) must be used, not the total 2DHG density ($p_{2d}$). To otherwise use $p_{2d}=p_1+p_2$ for calculating $E_{\textsc{f}}$ will overestimate it by a factor of $p_{2d}/2p_1$ [e.g., the error is 48\% at $p_{2d} = 1.9\times 10^{15}$/m$^2$ in Fig.\,\ref{fig:Basics}(c)].

It is interesting to note that in asymmetric quantum wells the experimental values of $m_1$ ($\approx$\,0.38$m_e$) also do not appear to vary significantly as a function of 2DHG density \cite{nichele2014spin}, from $p_{2d} = 1.3\times 10^{15}$/m$^2$ to $3.0\times 10^{15}$/m$^2$. This implies the parabolic nature of the HH$-$ subband dispersion is not only limited to single heterojunctions, but is a general feature of \textit{any} partially spin-orbit polarized and asymmetrically confined GaAs 2DHG.

As long as $m_1$ remains constant as a function of 2DHG density, the LH/HH anticrossing has not yet been reached [see Fig.\,\ref{fig:Basics}(a)].  As a 2DHG goes through the anticrossing, the second lowest occupied hole subband switches from HH$-$ to LH$+$, with the lowest occupied subband remaining HH$+$. Any mass measurement before/after the anticrossing would thus reveal a sudden step change in ``$m_1$'' (recall that $m^*_{\textsc{lh}} > m^*_{\textsc{hh}}$ in the 2DHG in-plane direction) but not in $m_2$. This new ``$m_1$'' (from LH$+$) would become density-dependent. As a simpler alternative to $m^*$ measurements, we propose to monitor $\Delta p/p$ instead for a signature of the LH/HH anticrossing. As the 2DHG goes through the anticrossing, one should see a maximum in $\Delta p/p$ as function of 2DHG density [see Fig.\,\ref{fig:Basics}(a)]. In any case, using either signature ($m_1$ discontinuity or $\Delta p/p$), our sample has clearly not yet reached the LH/HH anticrossing [see Fig.\,\ref{fig:Basics}(d) and Fig.\,\ref{fig:FFT}(b)]. A Landau fan measured at high magnetic fields (up to $B=12$~T; not shown) confirmed that no other subband (either the lowest LH subband or the second-lowest HH subband) is involved in transport.

The disorder $h\Gamma=h/\tau$ implied by the quantum scattering times $\tau_1$ and $\tau_2$ from Fig.~\ref{fig:FFT}(a) are consistent with the disorder-driven reduction of the Landau level spacing $(\hbar \omega_c-h\Gamma)$ where $\omega_c$ is the cyclotron frequency, measured at high magnetic fields (not shown).  The transport scattering time (a.k.a. momentum relaxation time) $\tau_{tr} = m^*_{\parallel} \mu_h/e$ (where $e$ is the electron elementary charge) implied by the Hall hole mobility $\mu_h$ is an order of magnitude larger than $\tau_{1,2}$.

Figure~\ref{fig:FFT}(b) shows that $m_2$ changes as a function of the 2DHG density: the dispersion of the HH$+$ subband is not parabolic. The only previously reported direct SdH measurement of $m_2$ in a GaAs (100) single heterojunction, $m_2= (0.8\pm0.1)m_e$ at $p_{2d} = 2.3\times 10^{15}$/m$^2$ \cite{habib2004spin}, agrees with that in sample H, $m_2= (0.81\pm0.06)m_e$ at $p_{2d} = 2.2\times 10^{15}$/m$^2$. Both follow the trend extrapolated from $m_2$ in sample~J.

Next, we compare our $m_1$ and $m_2$ values to cyclotron resonance (CR) mass measurements in (100) GaAs/AlGaAs single heterojunctions as a function of 2DHG density, using published CR studies that are the closest available in experimental conditions.

\begin{figure}[t]
    \includegraphics[width=1.0\columnwidth]{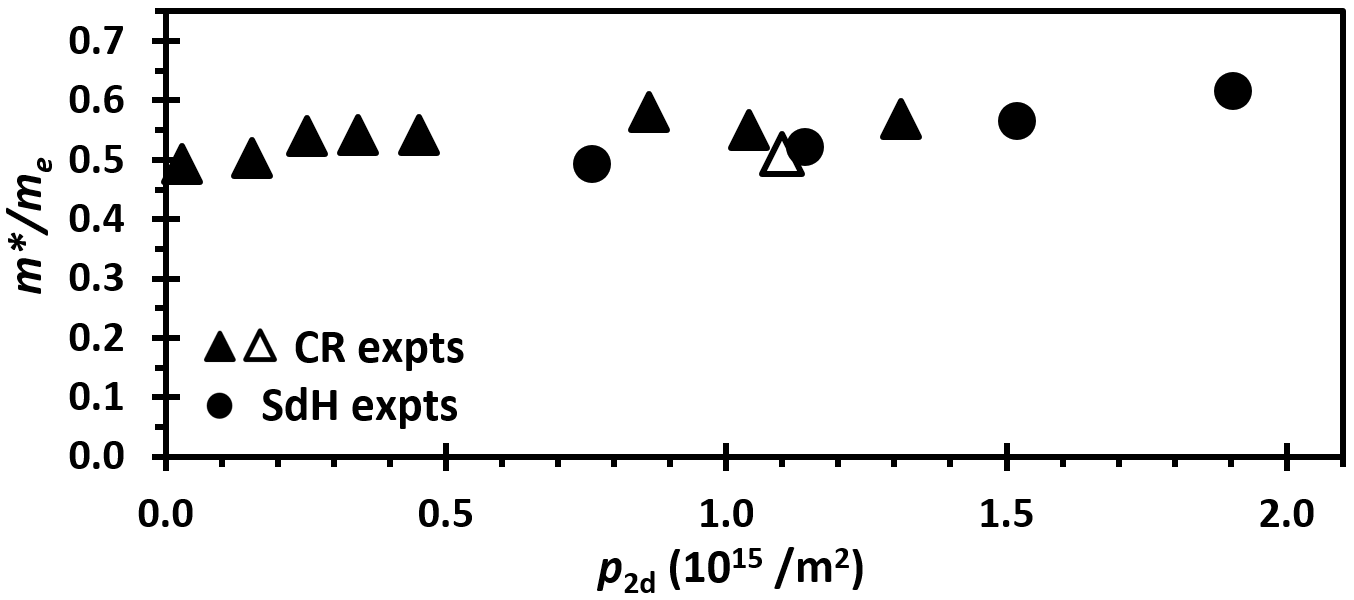}
    \caption{Comparison of 2DHG effective mass measurements from cyclotron resonance (CR) experiments and our SdH experiments in (100) GaAs single heterojunctions. Each filled (empty) triangle symbol represents a separate modulation-doped wafer from Ref.\,\onlinecite{lu2008cyclotron} (Ref.\,\onlinecite{zhu2007density}), measured at T\,=\,4\,K. Circles represent the weighted averages $(p_1m_1+p_2m_2)/(p_1+p_2)$ of the low-temperature data in sample~J, to emulate T\,=\,4\,K (see main text). For all data, error bars are similar in size or smaller than the symbols. }
    \label{fig:CRmass}
\end{figure}

At low temperature ($T$\,=\,0.5\,K), two CR peaks are typically observed \cite{stormer1983energy, Cole1997}, corresponding to $m_1$ and $m_2$ from the spin-orbit split HH$-$/HH$+$. The CR mass $m_1=0.38m_e$ agrees with the SdH mass $m_1=(0.36 \pm 0.03)m_e$ in the same sample at $p_{2d} = 5.0 \times 10^{15}$/m$^2$ \cite{stormer1983energy}, which also agrees with our SdH mass $\overline{m_1}=(0.35\pm 0.01)m_e$ at lower 2DHG densities. At high temperature ($T$\,=\,4\,K), only a single CR peak is typically observed \cite{Cole1997, zhu2007density, lu2008cyclotron}. However, it has been shown that the two CR peaks corresponding to $m_1$ and $m_2$ from spin-orbit split HH$-$/HH$+$ merge non-trivially into a single CR peak as temperature increases from $T$\,=\,0.5\,K to 4\,K \cite{Cole1997}. The frequency of this single, merged CR peak at $T$\,=\,4\,K is proportional to the weighted average $(p_1m_1+p_2m_2)/(p_1+p_2)$ of the CR peaks/masses observed at $T$\,=\,0.5\,K [see Eq.(5) in Ref.\,\onlinecite{Cole1997}]. Therefore we consider the single broadened CR peaks observed at $T$\,=\,4\,K for the many single heterojunctions in Refs.~\onlinecite{lu2008cyclotron, zhu2007density} to originate from the merging of two underlying CR peaks corresponding to $m_1$ and $m_2$.

Figure~\ref{fig:CRmass} shows that the weighted averages of our SdH $m_1$ and $m_2$ values (circles) are very close to the observed CR effective mass $\sim$\,0.5$m_e$ (triangles) reported in Refs.\,\onlinecite{zhu2007density,lu2008cyclotron}, and follow a similar slow upwards trend. Thus we conclude that experimental SdH masses agree with experimental CR masses in (100) single heterojunctions. Identifying general trends is much less straightforward for quantum wells than in single heterojunctions, as more variables compete, such as quantum well width, wavefunction symmetry, electric field, 2DHG density and mobility, barrier alloy composition, and strain.

We next compare the published CR masses of 2DHGs hosted in GaAs/AlGaAs single heterojunctions along different crystal plane orientations, (311)A and (100). At low temperature ($T$\,=\,0.5\,K), two CR peaks are observed, giving $m_1=0.21 m_e$ and $m_2=0.46 m_e$ in (311)A \cite{Cole1997}, much smaller than the $m_1=0.38 m_e$ and $m_2=0.60 m_e$ found in (100) \cite{stormer1983energy}. At high temperature ($T$\,=\,4\,K), a single CR peak is observed, giving $m^* \approx 0.37 m_e$ in (311)A \cite{Hirakawa1993,Cole1997}, again smaller than the $m^* \approx 0.5 m_e$ found in (100) \cite{zhu2007density,lu2008cyclotron}. As with (100), the (311)A CR masses at $T$\,=\,4\,K \cite{Hirakawa1993,Cole1997} agree with those at $T$\,=\,0.5\,K \cite{Cole1997}, according to $(p_1m_1+p_2m_2)/(p_1+p_2)$. Since $m_1$ does not vary with 2DHG density, the origin for $m_1$ in (311)A being smaller than those in (100) cannot be due to differences in 2DHG density. And since SdH and CR masses from single heterojunctions in similar experimental conditions are equivalent, we thus conclude that heavy hole effective masses of 2DHGs hosted in (311)A GaAs single heterojunctions are in general much smaller than those in (100).

\begin{figure}[t]
    \includegraphics[width=1.0\columnwidth]{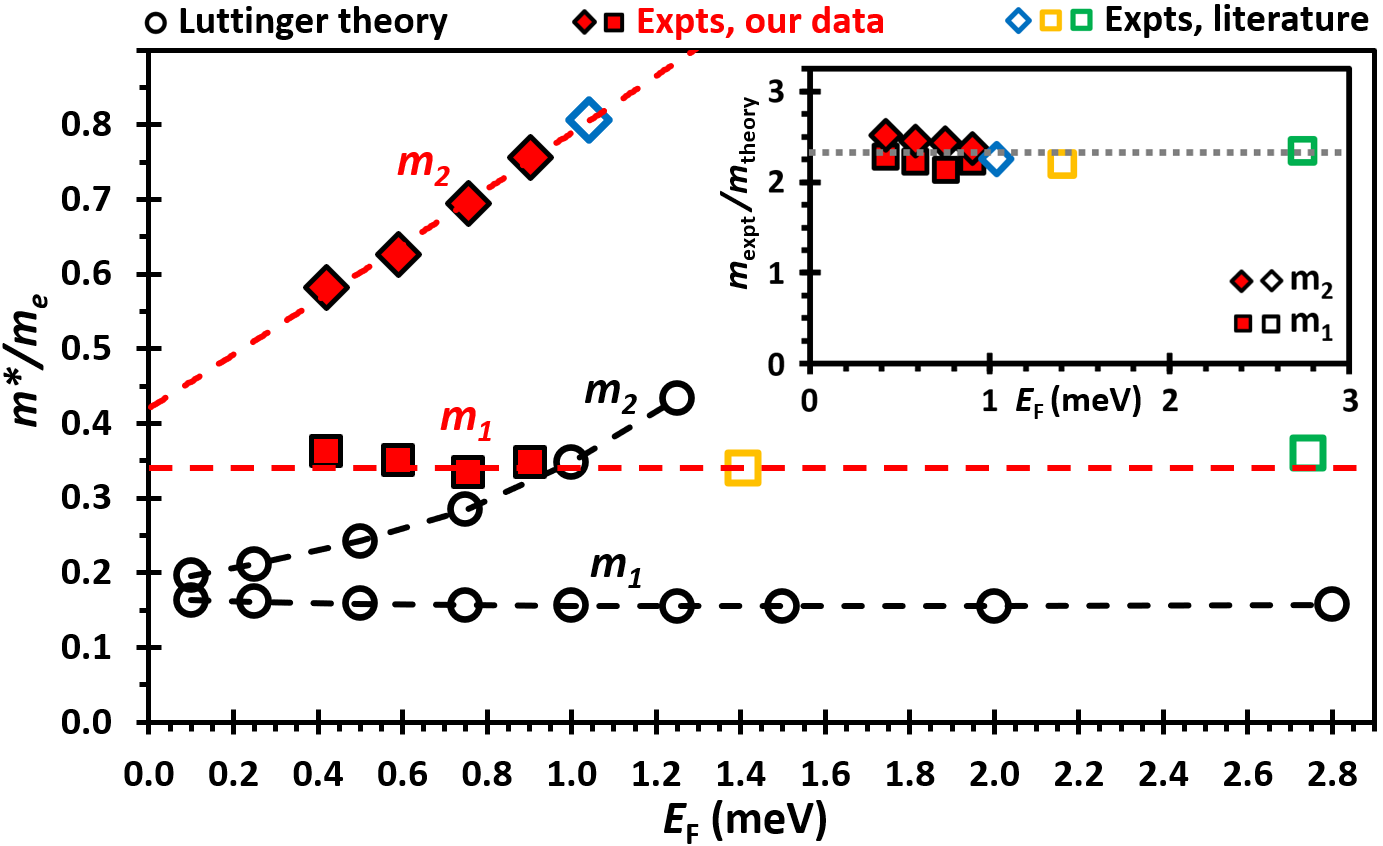}
    \caption{Comparison of $m_1$ and $m_2$ values obtained from our magnetotransport results in sample~J (full diamonds and squares), published SdH studies (empty diamonds and squares) \cite{stormer1983energy, habib2004spin, grbic2004}, and Luttinger theory (empty circles) \cite{companionPRB2025}. The inset demonstrates that all experimental $m_1$ (squares) and $m_2$ (diamonds) are larger than Luttinger theory predictions by approximately the same scaling factor ($\approx$\,2.3). All dashed and dotted lines are polynomial fits to the data. }
    \label{fig:Model}
\end{figure}

We now directly compare theoretical predictions to our magnetotransport results. Since the HH$-$ and HH$+$ subbands share the same Fermi energy [see Fig.\,\ref{fig:Basics}(a)] and since $E_{\textsc{f}}$ can now be accurately calculated from experiments with Eq.~(\ref{Eq:E_F(p1)}), Figure~\ref{fig:Model} overlays our experimental $m_1$ and $m_2$ as well as those from published SdH studies \cite{stormer1983energy, habib2004spin, grbic2004} as a function of $E_{\textsc{f}}$ with numerical calculations based on the standard Luttinger model \cite{Luttinger56}. These calculations are extensively described in Ref.~\onlinecite{companionPRB2025}.

The experimental results and theoretical predictions in Fig.\,\ref{fig:Model} bear three striking similarities. First, both reveal a near-independence of $m_1$ on the Fermi energy, supporting our claim that the HH$-$ subband is essentially parabolic for the investigated range of $E_{\textsc{f}}$ values (all below the anticrossing point between the HH$-$ and the LH$+$ subbands). Second, both reveal a simple near-linear/quadratic dependence of $m_2$ at low $E_{\textsc{f}}$ values (0.4$-$1.0~meV). Third, both show $m_2$ approaching $m_1$ as $E_{\textsc{f}}$ decreases, a behavior widely expected but never observed before in any 2D hole system.

Remarkably, the experimental values of \textit{both} masses $m_1$ and $m_2$ in  Fig.\,\ref{fig:Model} are larger than theory predictions by a common scaling factor ($\approx$\,2.3), and this common scaling factor remains approximately constant over the entire range of experimental Fermi energies (see inset of Fig.\,\ref{fig:Model}). Since we established earlier that our $m_1, m_2$ masses agree with those already reported in SdH and CR mass studies in (100) single heterojunctions, the disagreement between theory and experiment in Fig.\,\ref{fig:Model} also extends to those previous reports \cite{stormer1983energy, Cole1997, zhu2007density, lu2008cyclotron, habib2004spin, grbic2004}. The enhancement factor of $\sim$\,2.3 appears near-universal as it occurs in (100) single heterojunctions from five MBE research groups, in different types of experiments (CR and SdH), and over a very wide range of 2DHG densities from $p_{2d} = 0.7 \times 10^{15}$/m$^2$ to $5 \times 10^{15}$/m$^2$.

The Luttinger calculations in Fig.\,\ref{fig:Model} do not include many-body interactions, and use single-particle bulk GaAs band parameters obtained from optical studies \cite{Adachi1985GaAs, Pavesi1994photoluminescence, Winkler1995excitons, Vurgaftman2001, HerbertLi2001}. The strength of many-body interactions in 2D systems is typically characterized by the interaction parameter $r_s = m^*_{\textsc{hh}} e^2 / 4\pi \hbar^2 \epsilon \sqrt{\pi p_{2d}}$ \cite{Spivak2010}, where $\epsilon$ is the dielectric constant. By that metric \footnote{The relationship $r_s = m^*_{\textsc{hh}} e^2 / 4\pi \hbar^2 \epsilon \sqrt{\pi p_{2d}}$ is not strictly valid, since its derivation assumes a parabolic dispersion.}, all experimental data shown in Fig.\,\ref{fig:Model} (including previously published studies) is well into the strong hole-hole interaction regime with $r_s$ values ranging from 6 (at $E_{\textsc{f}}=2.8$~meV) to 16 (at $E_{\textsc{f}}=0.4$~meV), using the density-weighted average effective mass $(p_1m_1+p_2m_2)/(p_1+p_2) \approx 0.5 m_e$.

In the high-$r_s$ regime ($r_s > 3$), the effective mass is generally expected to be renormalized to larger values than in the non-interacting regime $r_s < 1$ \cite{Spivak2010}. For 2D electron gases (2DEG) in this regime, theory predicts the effective mass should be renormalized by a factor of 2$-$3 \cite{KwonY1994, Zhang2005, Asgari2005, Asgari2006}, and this has been experimentally observed in Si \cite{Pudalov2002, Shashkin2003}, GaAs \cite{Tan2005}, ZnO \cite{Kasahara2012}, and AlAs \cite{Vakili2004}. Since, as discussed in \cite{companionPRB2025}, the larger-than-predicted $m_1$ and $m_2$ in Fig.\,\ref{fig:Model} are most likely not due to magnetic field, strain, or incorrect Luttinger parameters (i.e., $\gamma_1, \gamma_2, \gamma_3$), we hypothesize they are indeed due to many-body interactions. Our many-body interaction hypothesis does not invalidate the equivalence we noted earlier between CR and SdH mass experiments. The Kohn theorem \cite{Kohn1961}, which states that electron-electron interactions do not contribute to effective masses measured in CR experiments, does not apply to non-parabolic subbands or to spin-split carriers with different masses \cite{Ando1982electronic,Cole1997,Peeters1990}.

When $p_{2d} < 2 \times 10^{15}$/m$^2$, the effective mass enhancement factor ($\sim$\,2.3) has a very weak density dependence: the inset of Fig.\,\ref{fig:Model} shows that the ratio $m_\text{expt}/m_\text{theory}$ rises slightly as $E_\textsc{f}$ decreases for both $m_1$ and $m_2$. For example, the ratio for $m_2$ at $E_\textsc{f} = 0.4$~meV is about 1.07 times the ratio at $E_\textsc{f} = 0.9$~meV. Figure~\ref{fig:FFT}(a) demonstrates a similar trend for $m_1$ in sample~H. That behavior is consistent with past suggestions that spin-orbit interactions specifically associated with total angular momentum $J=3/2$ and LH/HH hybridization compete with many-body interactions (see discussion in Sections IV.B and IV.C from \cite{winkler2005anomalous}), which causes the exchange energy to vary slowly rather than fast with 2DHG density (see Fig.~2 from \cite{Kernreiter2013}). This competition between $J=3/2$ spin-orbit and many-body interactions could explain why the 2DHG density dependence of the effective mass is much more subdued than that with electrons in the high $r_s$ regime.

\textit{Summary and outlook.} We demonstrated that the dispersion of the HH$-$ subband in (100) GaAs 2DHGs is very close to a parabola at Fermi energies below the LH/HH anticrossing, as evidenced by its density-independent effective mass ($m_1$). Although our experiments focused on single heterojunctions, these findings appear to more generally apply to any asymmetrically-confined GaAs 2DHG in quantum wells.

Our values for $m_1$ and $m_2$ agree with those found at a single 2DHG density in cyclotron resonance studies and low-field Shubnikov-de Haas magnetotransport studies with (100) GaAs/AlGaAs single heterojunctions.

Numerical calculations based on the Luttinger model matched the density dependence of our $m_1$ and $m_2$, but they underestimated both experimental values by a common factor ($\approx$\,2.3). We propose the difference is due to many-body interactions. The mass enhancement itself appears to be density-independent, possibly due to interplay with spin-orbit interactions. To accurately reproduce experimental results, future theory models should include many-body interactions while still predicting a parabolic dispersion for the HH$-$ subband at $E_\textsc{f}$ values below the LH$+$/HH$-$ anticrossing.

The combination of $J=3/2$ valence bands, strong structural inversion asymmetry, and large interaction parameter $r_s$ is common to many 2D holes systems. We therefore expect that a nearly-parabolic HH$-$ branch and a largely density-independent many-body enhancement of both spin-orbit-split masses will be features of such platforms, with the quantitative value of the enhancement factor remaining material- and structure-dependent.

\begin{acknowledgments}
J.B. and J.B.K. supervised this work equally. This research was undertaken thanks in part to funding from the Canada First Research Excellence Fund (Transformative Quantum Technologies), Defence Research and Development Canada (DRDC), National Research Council Canada (NRC), Canada's Natural Sciences and Engineering Research Council (NSERC), and the Quantum Sensors Challenge program from NRC. S.R.H. acknowledges further support from the NSERC Canada Graduate Scholarships $-$ Doctoral program. The University of Waterloo's QNFCF facility was used for this work. This infrastructure would not be possible without the significant contributions of CFREF-TQT, Canada Foundation for Innovation (CFI), Innovations, Science and Economic Development Canada (ISED), the Ontario Ministry of Research, Innovation and Science, and Mike and Ophelia Lazaridis. Their support is gratefully acknowledged.
\end{acknowledgments}

\end{document}